\documentclass[
extra
,mreferee
]{gji2}
\usepackage{graphicx}
\graphicspath{ {./figs/} } 
\usepackage{timet}
\usepackage{amsmath}
\usepackage{amsfonts}
\usepackage{bm}
\usepackage{color}
\usepackage{lineno}

\usepackage[urlcolor=blue,citecolor=black,linkcolor=black]{hyperref}

\title[Adjoint-based uncertainty quantification for inhomogeneous friction on a slow-slipping fault]
  {Adjoint-based uncertainty quantification for inhomogeneous friction on a slow-slipping fault}
\author[S. Ito et al.]
  {Shin-ichi Ito$^{1,2}$, Masayuki Kano$^3$, and Hiromichi Nagao$^{1,2}$ \\
  $^1$ Earthquake Research Institute, The University of Tokyo, Japan,\\
  $^2$ Graduate School of Information Science and Technology, The University of Tokyo, Japan,\\
  $^3$ Graduate School of Science, Tohoku University, Japan
  }
\date{\today}
\pagerange{\pageref{firstpage}--\pageref{lastpage}}
\volume{??}
\pubyear{??}

\newcommand{\rmd}{\mathrm{d}}
\newcommand{\rmx}{\bm x}
\newcommand{\rmz}{\bm z}
\newcommand{\rmb}{\bm b}

\newcommand{\dt}{\frac{\mathrm{d}}{\mathrm{d}t}}

\newcommand{\Y}{\bm Y}

\newcommand{\hesse}{\mathsf{H}}

\newcommand{\rmtl}{\bm \xi}
\newcommand{\rmadj}{\bm \lambda}
\newcommand{\rmsoa}{\bm \eta}

\newcommand{\rmy}{\bm y}
\newcommand{\obsoperator}{\bm H}
\newcommand{\rmw}{\bm w}
\newcommand{\rmf}{\bm f}
\newcommand{\rmh}{\bm h}

\newcommand{\rmv}{\bm v}

\newcommand{\real}{\mathbb{R}}
\newcommand{\dimx}{n}
\newcommand{\dimy}{m}
\newcommand{\dimsim}{d}

\newcommand{\sdd}[3]{\frac{{#1}^2{#2}}{{#1}{#3}{#1}{#3}^{\top}}}
\newcommand{\sddinline}[3]{{#1}^2{#2}\slash{#1}{#3}{#1}{#3}^{\top}}

\newcommand{\revised}[1]{\textcolor{black}{#1}}

\newcommand{\preprintfigsize}{0.95\linewidth}

\newcommand*\patchAmsMathEnvironmentForLineno[1]{
  \expandafter\let\csname old#1\expandafter\endcsname\csname #1\endcsname
  \expandafter\let\csname oldend#1\expandafter\endcsname\csname end#1\endcsname
  \renewenvironment{#1}
     {\linenomath\csname old#1\endcsname}
     {\csname oldend#1\endcsname\endlinenomath}}
\newcommand*\patchBothAmsMathEnvironmentsForLineno[1]{
  \patchAmsMathEnvironmentForLineno{#1}
  \patchAmsMathEnvironmentForLineno{#1*}}
\AtBeginDocument{
\patchBothAmsMathEnvironmentsForLineno{equation}
\patchBothAmsMathEnvironmentsForLineno{align}
\patchBothAmsMathEnvironmentsForLineno{flalign}
\patchBothAmsMathEnvironmentsForLineno{alignat}
\patchBothAmsMathEnvironmentsForLineno{gather}
\patchBothAmsMathEnvironmentsForLineno{multline}
}

\begin{document}

\maketitle

\begin{summary}

Long-term slow-slip events (LSSEs) usually occur on a fault existing at the deep, shallow parts of subducting plates and substantially relate to adjacent megathrust fault motions.
The dynamics of the LSSE largely depend on the inhomogeneity of friction that occurs between the fault interfaces.
Thus, it is crucial to estimate the spatial-dependent frictional features from the observations of the slip motion and subsequently identify essential parts that contribute to the principal slip motion by quantifying uncertainties involved in the estimates.
Although quantifying the uncertainties of the frictional feature fields in high-resolution is necessary to solve the task, conventional techniques of quantifying slow earthquake frictional features have not yet achieved such uncertainty quantification (UQ) due to the complexity of LSSE models such as the large dimensionality.
We, therefore, propose a method of UQ for spatially inhomogeneous frictional features from slip motion based on a four-dimensional variational data assimilation technique using a second-order adjoint method.
The proposed method enables us to conduct an accurate UQ even when the dimensionality is large.
By combining a fault motion model that mimics slow-slip motion on an LSSE fault––megathrust fault complex in southwestern Japan and the data assimilation technique, we successfully quantified the spatial distribution of the uncertainty of the frictional features in high-resolution.
The evaluated spatial distribution in high-resolution reveals the correlation between the dynamics of the slow-slip motion and the important components of the frictional features, which is valuable information for design of observation systems. Findings from this study are expected to advance the theoretical foundation of applied seismic motion prediction techniques using slow-slip frictional features as stress meters for megaquakes, as well as improve understanding of the relationship between the slow-slip motion and frictional parameters of a fault.

\end{summary}


\section{Introduction}\label{intro}

Slow earthquakes have received vast attention as stress meters for megaquakes (\cite{Ob16}) and have been detected worldwide (\cite{ka18}) since the first discovery of the Bungo Channel long-term slow-slip event (LSSE) in 1997 on the Philippine Sea Plate interface located in southwestern Japan (\cite{Hi99,Ob02}), which is referred to as the Bungo Channel LSSE fault.
Slow earthquakes usually occur on the deeper and shallower parts of subducting plates, and such activity largely affects adjacent megathrust faults.
Therefore, understanding the physical properties of slow earthquakes is important for obtaining evidence of potential related megaquakes that may occur in the future.
Since the dynamics of the slip motion of a fault largely depend on the spatially inhomogeneous frictional features of the fault, it is essential to estimate the spatial dependence of the frictional features from the observations of the slip motion and to identify essential components of the frictional features that contribute to the main motion.
In recent years, data assimilation (DA) has received increasing attention as a method for estimating such frictional features.
The DA method is a statistical technique to integrate numerical simulation models and observational data based on Bayesian statistics, which was originally developed in the fields of meteorology and oceanography (\cite{ka03,TSU07}) and enables researchers to estimate unobservable quantities such as the spatial frictional features of faults.
Data assimilation methods can be classified into two types: sequential DA and non-sequential DA.
Based on the ensemble Kalman filter (\cite{Ev03}), classified under sequential DA, \cite{Hi19} has conducted numerical experiments to estimate the frictional features on a fault along the Bungo Channel by combining observations of the slip motion on the fault and a fault motion model. The established fault motion model, which is explained in Section \ref{model}, is applied for the investigation of the LSSE motion on the Bungo Channel LSSE fault. 
In other slip motion research, \cite{ka15} has applied a four-dimensional variational (4DVar) method (\cite{Le86,Pl06}) classified under non-sequential DA (herein, the 4DVar DA method) to estimate the frictional features on a fault producing afterslip, which is also a slow-slip motion occurring after the earthquake, from post-seismic crustal deformation data. The 4DVar DA method can therefore be a suitable approach to evaluate fault slip motion utilizing DA.
Although DA has been used for obtaining basic estimates of the frictional fault features in these studies, in theory, DA can provide even richer information, such as the uncertainty involved in the estimates.
This is because, mathematically, the aim of DA is to evaluate a specific probability density function (PDF) called the posterior PDF, which is a conditional PDF of stochastic variables involved in a numerical model with a given observational dataset.
Since the posterior PDF is defined in a high dimension that is equal to the number of variables in the numerical model, obtaining the uncertainty from the posterior PDF becomes extremely difficult when using a large-dimensional model.
For this reason, the previous studies have assumed the coarse-grained spatial distribution of the frictional features to suppress the increase in the number of variables to be estimated.
However, evaluating the spatial distribution of the uncertainty of the frictional features in high-resolution is critical for understanding the relationship between the physical properties of the slip motion and the frictional features themselves.
Discussing such a relationship based only on the coarse-grained results is inherently difficult, and the uncertainty quantification (UQ) of the spatial distribution of the frictional features in high-resolution is essential.
Considering this, a recent development to the 4DVar DA method by \cite{It16} based on a second-order adjoint (SOA) method (\cite{wa98,Di02}), hereinafter the SOA-based UQ method, has enabled us to obtain estimates together with their uncertainties within a practical computational timeframe using an appropriate amount of resources.
The SOA-based UQ method was originally developed to analyze the uncertainty of physical parameters and/or initial conditions involved in a large-scale model of crystal growth in a metal alloy (\cite{It17}); however, it can also be applied to all other numerical models since its mathematical framework is general.
Therefore, application of the SOA-based UQ method would enable us to obtain for the first time the spatial dependence of the uncertainty of frictional features in high-resolution.
To this end, this study developed a method to (1) quantify the spatial distribution of the uncertainty of frictional features based on \cite{It16}, and (2) elucidate the relations between them and the slip motion.
Quantifying the frictional feature–-slip motion relationship is expected to provide valuable information related to observational design, e.g., timing and duration of data acquisition.
In addition, the numerical simulations of the LSSE, considering the uncertainty of the frictional features, will allow the prediction of crustal deformation and its uncertainty.
The remainder of this paper is organized as follows:
Section~\ref{model} introduces the fault motion model employed in this study and observes the kinetic characteristics of the model through its numerical simulation.
Section~\ref{DA} proposes the SOA-based UQ method to be applied to the fault motion model.
The proposed method enables the quantification of the uncertainty within a more realistic computational timeframe and reasonable use of resources even when the scale of the model is large.
Section~\ref{App} describes the application to the fault motion model and then presents the numerical experiments that were conducted to quantify the uncertainties involved in the frictional parameters.
The experiments reveal how the uncertainties are affected by the activities of the slip motion and/or the amount and quality of data to be assimilated.
Section~\ref{conclusion} concludes this work, including a discussion of certain limitations and relevant future research directions.

\section{model}\label{model}
\begin{figure*}
 \centering
 \includegraphics[width=0.99\linewidth]{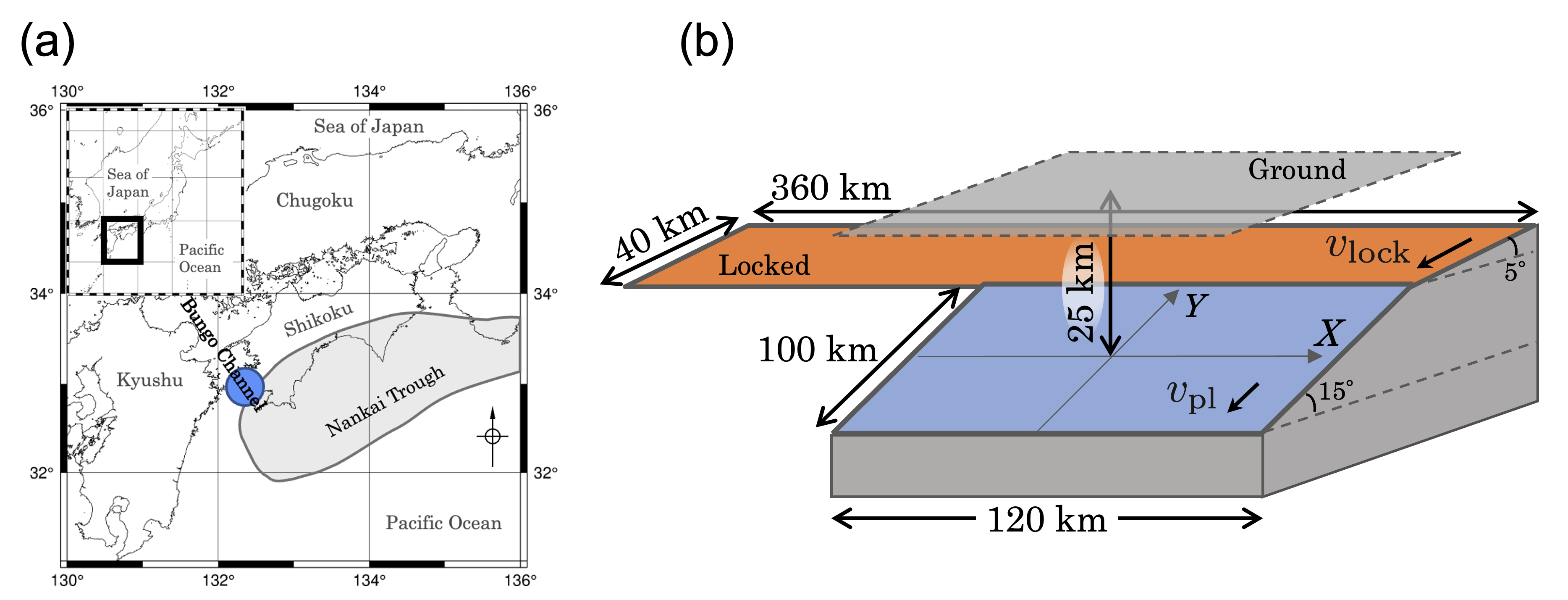}
 \caption{(a): Map of the western part of southern Japan, showing the Bungo Channel region which was investigated in this study. (b): Schematic view of the discrete model proposed by \cite{Hi19}.}
 \label{geom}
\end{figure*}

\begin{figure*}
 \centering
 \includegraphics[width=0.99\textwidth]{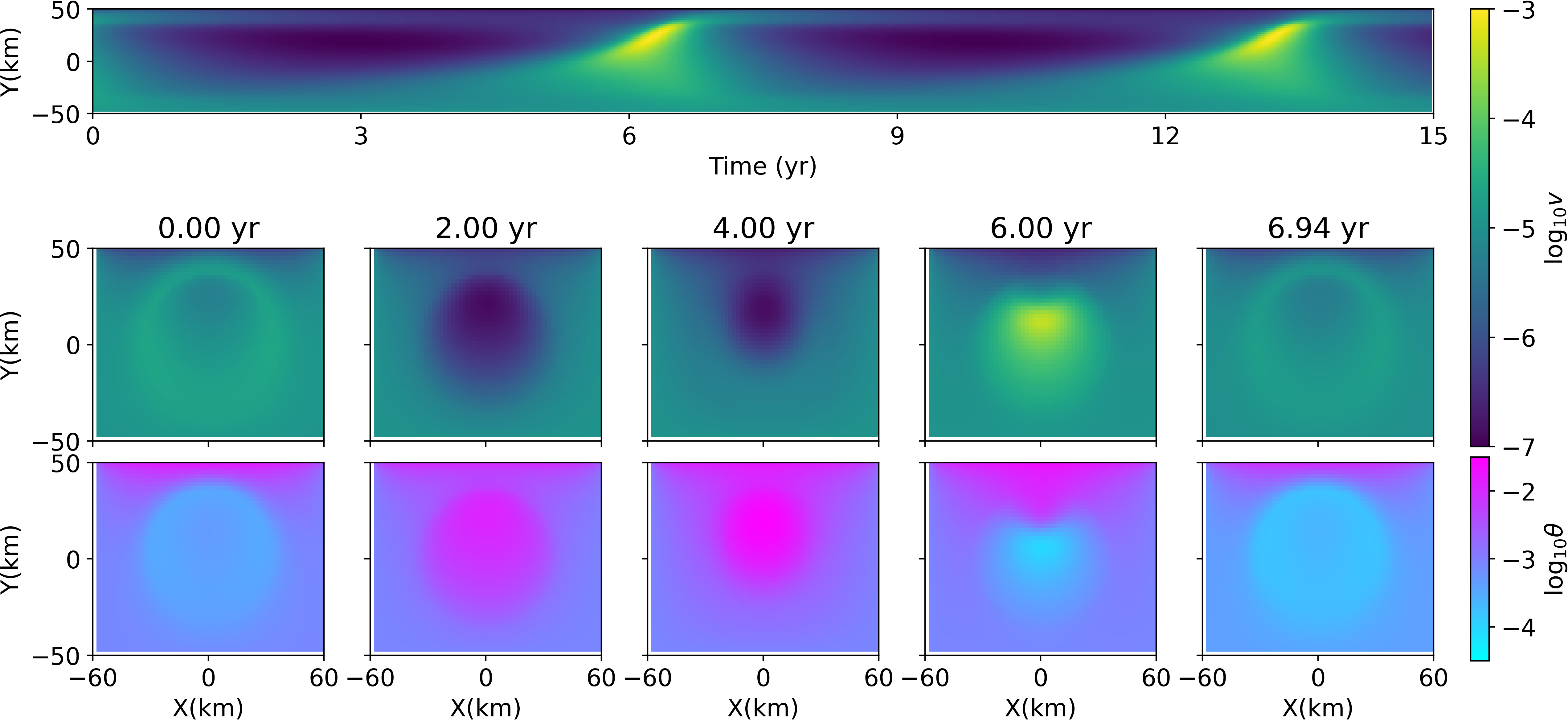}
 \caption{Time evolution of the velocity $v$ and state variable $\theta$ fields using the set of parameters shown in Table~1.
 The units of $v$ and $\theta$ are $\;\text{m}\slash\text{yr}$ and $\text{yr}$, respectively. The top panel shows the time evolution of $v$ on $X=0$ (km) line.}
 \label{evolution}
\end{figure*}

This paper employs the discrete fault motion model called “Bungo Channel LSSE Model 2” proposed by \cite{Hi19} to describe an LSSE motion on the Bungo Channel LSSE fault (see Section 3.2.2 of that paper).
The Bungo Channel, with its associated LSSE fault, located in southwestern Japan (see Fig.~\ref{geom}(a)), is known as a location at which the LSSEs on the main fault are observed with slip durations of 1 year and have been occurring with a recurrence time of 6--7 years (\cite{Yo15}).
The updip portion of the Bungo Channel LSSE fault features a megathrust fault that periodically causes events known as the Nankai megaquakes, with intervals of several tens or several hundreds of years (\cite{An75}) and which interact with the Bungo Channel LSSE.
The Bungo Channel LSSE Model 2 considers these two faults (the Bungo Channel LSSE fault and the megathrust fault) as having different time scales of the slip activities driven by an associated plate subducting at a steady rate $v_{\text{pl}}$, as shown in Fig.~\ref{geom}(b).
The blue rectangular region illustrates the Bungo Channel LSSE fault, which is of interest in this study, and the orange region termed “Locked” illustrates the megathrust fault in the updip portion of the Bungo Channel.
The slip motion on the blue rectangular region is characterized by a slip field along the plate convergence direction.
The model discretizes the region by a set of small square-cell faults with a side length of $2\;\text{km}$.
The slip field $u_{i}(t)$ $(i=1,\dots,d)$ is defined on each cell, where $d$ is the number of the cells and $t$ is time.
The slip motions on the cells influence each other through long-range elastic interaction, which is calculated as a Green’s function assuming the orange and blue faults are embedded in a homogeneous Poisson solid defined in a three-dimensional half-space (\cite{Ok92}).
The model considers not only the interactions between the cells but also the long-range interaction resulting from the slow activity of the megathrust fault.
Compiling these effects, the shear stress $\tau_{i}(t)$ $(i=1,\dots,d)$ of each cell along the plate convergence direction is modeled by
\begin{multline}\label{eom}
	\tau_{i}(t) = \sum_{j=1}^{d} K_{ij} \left( u_{i}(t)-v_{\text{pl}}t\right)
		+ k_{i}\left(v_{\text{lock}}-v_{\text{pl}}\right)t - \frac{G}{2c}v_{i}(t) \\
		{} \left(i=1,\dots,\dimsim\right),
\end{multline}
where $v_{i}(t)$ is the velocity along the plate convergence direction, defined by
\begin{equation}\label{eqvel}
	\dt u_{i}(t) = v_{i}(t) \quad
	\left(i=1,\dots,\dimsim\right).
\end{equation}
The first term on the right-hand side of Eq.~\eqref{eom} represents the long-range interaction from the slips of the other cells relative to the plate slip $v_{\text{pl}}t$. 
The matrix $K_{ij}$ $(i,j=1,\dots,d)$ describes the interaction through the Green’s function, which physically means the change in the static stress of the cell $i$, owing to a unit slip of the cell $j$.
The second term is the interaction from the megathrust fault to the cell $i$, where the coefficient $k_{i}$ $(i=1,\dots,d)$ is computed by the Green’s function as well as $K_{ij}$.
\revised{The constant $v_{\text{lock}}$ is the convergence velocity of the megathrust fault.}
The third term indicates the radiation effect~(\cite{Ri93}), where $G$ and $c$ are the shear modulus and sound speed of the transverse wave, respectively.
The shear stress balances with the friction force that occurs between the fault and the plate.
The friction force is characterized by the friction coefficient $\mu_{i}$ $(i=1,\dots,d)$, defined by
\begin{equation}\label{eqfric}
	\tau_{i}(t) = \mu_{i}(t) N_{i} \quad
	\left(i=1,\dots,\dimsim\right),
\end{equation}
where $N_{i}$ is an effective normal stress affecting at the cell $i$.
In the model, the normal stress is assumed to depend on the place but not on time.
For the time-dependent friction coefficient $\mu_{i}$, the model employs a rate-and-state friction law~(\cite{Di79}) given by 
\begin{equation}\label{eqrsf}
	\mu_{i}(t) = \mu^{\text{o}} + a_{i}\log\frac{v_{i}(t)}{v^{\text{o}}}
		 + b_{i}\log\frac{\theta_{i}(t)}{\theta^{\text{o}}} \quad
	\left(i=1,\dots,\dimsim\right),
\end{equation}
where $\mu^{\text{o}}$ is a reference frictional coefficient that satisfies $\mu_{i}(t) = \mu^{\text{o}}$ when the velocity $v_{i}(t)$ and the state variable $\theta_{i}(t)$ are given by their references $v^{\text{o}}$ and $\theta^{\text{o}}$, respectively.
The state variable $\theta_{i}(t)$ is assumed to obey an aging law~(\cite{Ru83}):
\begin{equation}\label{eqth}
	\dt \theta_{i}(t) = 1 - \frac{v_{i}(t)\theta_{i}(t)}{L_{i}}\quad
	\left(i=1,\dots,\dimsim\right).
\end{equation}
The spatially dependent parameter fields $a_{i}$, $b_{i}$ and $L_{i}$, determined by the physical properties of the media composing the fault, are one of the most important factors in this model to describe the slip motion.
Especially, the difference between the parameters $a$ and $b$ directly relates to the stick-slip motion.
Supposing the steady state of Eq.~\eqref{eqth} and then eliminating $\theta$ from Eq.~\eqref{eqrsf} yields the friction coefficient as
\begin{equation}\label{eqrsf2}
	\mu_{i}(t) = (a_{i}-b_{i})\log v_{i}(t) + \text{const},
\end{equation}
which states that the local stability of the system depends on the sign of $a_{i}-b_{i}$, and expresses that the spatial dependency of parameter fields are essential for the complex behavior of the slip motion on the fault.
In \cite{Hi19}, Eqs.~\eqref{eom}--\eqref{eqth} are integrated to a time evolution equation of $v_{i}(t)$ for computational convenience.
The equation is as follows:
\begin{multline}\label{eqdbk}
    \dt v_{i}(t) = \left(\frac{A_{i}}{v_{i}(t)} + \frac{G}{2c}\right)^{-1}
    \left\{
    \sum_{j=1}^{\dimsim}K_{ij}\left(v_{j}(t)-v_{\text{pl}}\right) \right.\\
    {} \left. + k_{i}\left(v_{\text{lock}}-v_{\text{pl}}\right)
    - \frac{A_{i}-\left(A_{i}-B_{i}\right)}{\theta_{i}(t)}
    \left( 1 - \frac{v_{i}(t)\theta_{i}(t)}{L_{i}} \right)
    \right\} \\
    {}  \left(i=1,\dots,\dimsim\right),
\end{multline}
where $A_{i} = N_{i}a_{i}$ and $B_{i} = N_{i}b_{i}$.
By solving this equation simultaneously with Eq.~\eqref{eqth}, we can obtain the slip motion on the fault of interest.
In the following, we observe the typical slip motion based on the parameter setup shown in \cite{Hi19}.
We set $v_{\text{pl}}=6.5\;\text{cm}\slash \text{yr}$, \revised{$v_{\text{lock}}=0.5\;\text{cm}\slash \text{yr}$,} $G=40\;\text{GPa}$, and $c=3\;\text{km}\slash \text{s}$.
The frictional parameters $A_{i}$ and $L_{i}$ are set to $100\;\text{kPa}$ and $2.2\;\text{mm}$, and these values are common in all of the cells.
The parameter $B_{i}$ is set to have two different parameter regimes: One is $135\;\text{kPa}$ within $35\;\text{km}$ from the center of the fault region, and $30\;\text{kPa}$ otherwise.
This inhomogeneity of $B_{i}$ causes the stick-slip motion on the fault due to the coexistence of the regions having different signs of $A_{i}-B_{i}$.
These parameters are summarized in Table~1.
\begin{table}\label{paramtable}
\caption{Parameter set used in Fig.~\ref{evolution}.}
\centering
  \begin{tabular}{|c|c|} \hline\hline
    Parameters & Values \\ \hline
    $d$ & $3,000$ ($60$ and $50$ cells for $X$- and $Y$-directions) \\ \hline
    $A_{i}$ & $1.0\times 10^{2}\;\text{kPa}$ \\ \hline
    $A_{i}-B_{i}$ & \begin{tabular}{rl}$-35 \;\text{kPa}$ & for \; $X^2+Y^2 \le \left( 35\;\text{km}\right)^2$ \\ $70 \;\text{kPa}$ & otherwise \end{tabular}
                  \\ \hline
    $L_{i}$ & $2.2\;\text{mm}$ \\ \hline
    $v_{\text{lock}}$ & $0.5 \;\text{cm}\slash \text{yr}$ \\ \hline
    $v_{\text{pl}}$ & $6.5 \;\text{cm}\slash\text{yr}$ \\ \hline
    $G$ & $40 \;\text{GPa}$ \\ \hline
    $c$ & $3.0\;\text{km}\slash\text{s}$ \\ \hline
\end{tabular}
\end{table}
The initial condition of the velocity and state variable are set to uniformly random variables in our simulation.
This model is believed to converge to a limited cyclic behavior of the stick-slip motion and is robust to the perturbation of the initial condition to some extent.
Our experiments confirm that the motion of the velocity and state variable converge to a periodic motion with a constant recurrence. 
With the parameter setup, we solve Eqs.~\eqref{eqdbk} and \eqref{eqth} numerically via the Runge--Kutta--Fehlberg (RKF45) time integrator with an adaptive step size control.
Figure~\ref{evolution} shows the time evolution of the velocity $v_{i}(t)$ and the state variable $\theta_{i}(t)$.
As seen in the cross-section view of the velocity field (the top panel of Fig.~\ref{evolution}), the slip motion occurs periodically with a recurrence of approximately 7 years.
In addition, the observation states that the slip motion is more active in the unstable region of the parameter fields than that of the other. 
The existence of the stable region is essential to reproduce the stick-slip motion of the fault since that region suppresses the spread of slipping.
Because of this fact, analyzing the detail of the spatial dependency of the parameter fields is important to understand the complexity of the slip motion of the fault.
However, owing to the measurement techniques available to date, it is impractical and nearly impossible to measure the spatial dependency of parameters on a fault that exists deep underground.
Therefore, the next objective of this study was to detect the components of the parameter fields that mainly contribute to the slip motion from the observational data.
To achieve such detection, the uncertainty of the parameter fields needs to be investigated with respect to the observed data of the slip motion. This is a challenging inverse problem as the computational complexity increases exponentially with the spatial resolution of the model, as mentioned in Section~\ref{DA}.
In fact, \cite{Hi19} constrained the resolution of the parameter fields by assuming the fields to be uniform or considering a set of a few patches.
Since numerical simulation models in seismology are becoming increasingly large-scale day-by-day, the reduction of the computational complexity in the UQ has become a target for improving the understanding of seismic properties.
Thus we recall the development and application of the SOA-based UQ method by \cite{It16,It17}, originally developed from the 4DVar DA method to analyze a reaction--diffusion model of crystal growth, which has provided a statistical method to rapidly quantify uncertainty.
As previously mentioned, we expect that the SOA-based UQ method will also work well when applied to seismic models since it accepts any models described by autonomous systems, in that the method would enable us to target the extraction of only the uncertainty components of interest within the linear computational complexity of the model resolution.
Therefore, the contribution of this study is the first application of our proposed DA method to the seismic model explained above, which utilizes the SOA-based UQ method to uncover the relationship between the main slip motion and the parameter fields through the quantification of the uncertainty.

\section{Data assimilation}\label{DA}
In this section, we provide a general formulation of DA, and then explain how each of the commonly used DA algorithms, including the 4DVar DA method, evaluate the posterior PDF and the algorithms’ required computational costs.
Thereafter, we introduce the SOA-based UQ method proposed by \cite{It16} and discuss its computational complexity.

\subsection{Four-dimensional variational method}
Let $\rmx_{t}$ be a $\dimx$-dimensional time-dependent state vector that follows an autonomous model given by an ordinary differential equation:
\begin{equation}\label{fwdeq}
    \frac{\rmd \rmx_{t}}{\rmd t} = \rmf\left(\rmx_{t}\right),
\end{equation}
where the function $\rmf:\mathbb{R}^{\dimx}\rightarrow\mathbb{R}^{\dimx}$ is assumed to be a second-order differentiable.
Note that the initial condition $\rmx_{t_{0}}$ is the unique control variable of the time evolution of $\rmx_{t}$.
In addition, suppose we have a large vector composed of $k$ observational data vectors  $\Y_{0:k}=\left(\rmy_{t_{0}}^{\top},\rmy_{t_{1}}^{\top},\dots,\rmy_{t_{k}}^{\top}\right)^{\top}$, where the subscript of $\Y_{0:k}$ means the set of the data from $t=t_{0}$ to $t=t_{k}$.
The symbol $\bullet^{\top}$ indicates the transpose of the quantity $\bullet$.
Each observation $\rmy_{t}\in\real^{\dimy}$ is assumed to have a relation
\begin{equation}\label{obsmodel}
\rmy_{t}  = \rmh( \rmx_{t}) + \rmw_{t}, 
\end{equation}
where $\rmh:\mathbb{R}^{\dimx}\rightarrow\mathbb{R}^{\dimy}$ is an “observation operator” that maps $\rmx_{t}$ to a quantity comparable with the data $\rmy_{t}$, and $\rmw_{t}\in\real^{\dimy}$ is mean-zero white noise vector that follows a PDF $q(\rmw)$.
Based on the given model and observational data, DA builds a conditional PDF $p\left(\rmx_{t}\mid\Y_{0:k}\right)$ termed, as mentioned earlier, as the posterior PDF.
In the case of the posterior PDF at $t=t_{0}$, i.e., $p\left(\rmx_{t_{0}}\mid\Y_{0:k}\right)$, it is given by the Bayes' theorem as 
\begin{equation}
	p\left(\rmx_{t_{0}}\mid\Y_{0:k}\right) = \frac{p\left(\rmx_{t_{0}}\right)p\left(\Y_{0:k}\mid\rmx_{t_{0}}\right)}{p\left(\Y_{0:k}\right)},
\end{equation}
where $p\left(\rmx_{t_{0}}\right)$ is a prior PDF that describes {\it a priori} knowledge of $\rmx_{t_{0}}$, and $p\left(\Y_{0:k}\mid\rmx_{t_{0}}\right)$ is a likelihood function that provides a relation between $\rmx_{t_{0}}$ and the observational data $\Y_{0:k}$.
The denominator on the right-hand side
\begin{equation}\label{norm}
p\left(\Y_{0:k}\right) = \int d\rmx_{t_{0}}\; p\left(\rmx_{t_{0}}\right)p\left(\Y_{0:k}\mid\rmx_{t_{0}}\right)
\end{equation}
is a normalization constant, which does not depend on $\rmx_{t_{0}}$.
According to the assumption of the observational data, the posterior density function is given by
\begin{equation}\label{post}
	p\left(\rmx_{t_{0}}\mid\Y_{0:k}\right) = \dfrac{p\left(\rmx_{t_{0}}\right)}{p\left(\Y_{0:k}\right)}\prod_{j=0}^{k} q\left(\rmy_{t_{j}}-\rmh(\rmx_{t_{j}})\right).
\end{equation}

In this context, the aim of DA would be to evaluate such a posterior PDF by extracting the relevant statistics.
Recall that the two methods of DA to evaluate $p\left(\rmx_{t}\mid\Y_{0:k}\right)$ are classified as sequential and non-sequential, as mentioned in Section~\ref{intro}.
Sequential DA, which is typified by an ensemble Kalman filter~(\cite{Ev03}) and particle filter~(\cite{Ki96,Na12}), computes $p\left(\rmx_{t_{i}}\mid \Y_{0:i}\right)$ $(i=0,\dots,k)$ sequentially by tracking the flow of a group of “particles” to construct a histogram approximation.
To obtain $p\left(\rmx_{t_{0}}\mid \Y_{0:k}\right)$, the sequential DA needs the computations going back in time from $p\left(\rmx_{t_{k}}\mid \Y_{0:k}\right)$ by using the memorized trajectories of particles; for this, an exceptionally large memory size is required.
In general, to obtain accurate statistics of $p\left(\rmx_{t_{0}}\mid \Y_{0:k}\right)$, the required number of particles reaches a quadratic order of $n$ for EnKF and an exponential order of $n$ for the particle filter.
It can be expected that the evaluation of $p\left(\rmx_{t_{k}}\mid \Y_{0:k}\right)$ would not be successful when the number of particles is insufficient; therefore, close attention should be paid to the obtained results and the robustness should be confirmed by changing the number of particles through multiple trial-and-errors (e.g., \cite{Sa18}).
Letting $C$ be the computations to solve Eq.~\eqref{fwdeq}, the computational complexity to obtain $p\left(\rmx_{t_{0}}\mid \Y_{0:k}\right)$ is proportional to a product of $C$ and the number of particles.
As the alternate approach, non-sequential DA, which is typified by the 4DVar DA method~(\cite{Le86}), generally requires fewer computations than sequential DA since it does not evaluate the full form of $p\left(\rmx_{t_{0}}\mid \Y_{0:k}\right)$.
The 4DVar DA method aims to obtain only an optimal $\hat{\rmx}_{t_{0}}$ for $\rmx_{t_{0}}$ that maximizes $p\left(\rmx_{t_{0}}\mid \Y_{0:k}\right)$.
For numerical convenience, the 4DVar DA method considers a minimization problem of the negative logarithmic of $p\left(\rmx_{t_{0}}\mid \Y_{0:k}\right) p(\Y_{0:k})$ given by
\begin{equation}
	J = \underbrace{-\log p\left(\rmx_{t_{0}}\right)}_{=I} +\sum_{j=0}^{k}\underbrace{\left[-\log q\left(\rmy_{t_{j}}-h(\rmx_{t_{j}})\right)\right]}_{=\mathcal{J}_{j}},
\end{equation}
with respect to $\rmx_{t_{0}}$, which is equivalent to the maximization of $p\left(\rmx_{t_{0}}\mid \Y_{0:k}\right)$. The minimization problem is usually solved by a gradient-based optimization method such as the steepest descent method, conjugate gradient method, and Broyden--Fletcher--Goldfarb--Shanno (BFGS) method (\cite{Br69,No80}).
Although the gradient of $J$ is needed in such a gradient-based optimization method, in the above definition of $J$, note that the term $I$ is directly differentiable with respect to $\rmx_{t_{0}}$, but the term $\mathcal{J}_{j}$ is not, since it depends on $\rmx_{t_{0}}$ implicitly through Eq.~\eqref{fwdeq}.
To obtain the gradient $\sum_{j=0}^{k} \rmd \mathcal{J}_{j}\slash \rmd \rmx_{t_{0}}$, the 4DVar DA method applies a variational approach to an augmented Lagrangian function defined by 
\begin{equation}\label{lag}
\begin{aligned}
\mathcal{L} & = \sum_{j=0}^{k}\mathcal{J}_{j} + \int_{t_{0}}^{t_{f}} dt\; \rmadj_{t}^{\top}\left(\rmf\left(\rmx_{t}\right)-\frac{\rmd \rmx_{t}}{\rmd t}\right) \\
& = \int_{t_{0}}^{t_{f}} dt\; \left[ \sum_{j=0}^{k} \delta\left(t-t_{j}\right)\mathcal{J}_{j} + \rmadj_{t}^{\top}\left(\rmf\left(\rmx_{t}\right)-\frac{\rmd \rmx_{t}}{\rmd t}\right) \right]
\end{aligned}
\end{equation}
where $t_{f}$ is a time later than $t_{k}$ and $\delta(t)$ is a Dirac delta function and $\rmadj_{t}$ is a Lagrange multiplier.
The calculus of variation of the Lagrangian function provides us a time evolution equation of $\rmadj_{t}$:
\begin{equation}\label{adjeq}
-\frac{\rmd \rmadj_{t}}{\rmd t} = \left(\frac{\partial\rmf}{\partial\rmx_{t}}\right)^{\top}\rmadj_{t} +\sum_{j=0}^{k} \delta\left(t-t_{j}\right)\frac{\partial\mathcal{J}_{j}}{\partial\rmx_{t_{j}}},
\end{equation}
together with the initial and end-point conditions
\begin{equation}\label{condadjeq}
\rmadj_{t_{0}} = \sum_{j=0}^{k}\frac{\rmd \mathcal{J}_{j}}{\rmd \rmx_{t_{0}}}, \quad  \rmadj_{t_{f}} = 0.
\end{equation}
It should be noted that Eq.~\eqref{adjeq} is termed as an adjoint model in the 4DVar DA methodology.
See \cite{Le86,wa92} for the details of the derivation of Eqs.~\eqref{adjeq} and \eqref{condadjeq} from Eq.~\eqref{lag}.
Based on this, the precise procedures to obtain the gradient $\rmd J\slash \rmd \rmx_{t_{0}}$ at a certain $\rmx_{t_{0}}$ are as follows:
First, we compute the trajectory of $\rmx_{t}$ from the given $\rmx_{t_{0}}$.
After that, we solve the adjoint model backwards in time starting from $t=t_{f}$ to $t=t_{0}$.
The influences of the observational data are integrated to $\rmadj_{t}$ through the delta function terms in Eq.~\eqref{adjeq}, which describe the misfit between the solution of the forward model and observational data.
Note that the partial derivatives $\partial\mathcal{J}_{j}\slash\partial\rmx_{t_{j}}$ in the delta function terms can be explicitly computed by substituting $\rmx_{t}$ with the definition. The details are explained in Section~\ref{App}, where the concrete form of $\mathcal{J}_{j}$, i.e., the definitions of $q(\rmw)$ and $\rmh( \rmx_{t})$, are provided.
Finally, we obtain the objective gradient by 
\begin{equation}\label{gradient}
\frac{\rmd J}{\rmd\rmx_{t_{0}}} = \frac{\rmd I}{\rmd \rmx_{t_{0}}} + \rmadj_{t_{0}}.
\end{equation}
The concrete form of $\rmd I\slash \rmd \rmx_{t_{0}}$ is shown in Section~\ref{App}, where the functional form of $p\left(\rmx_{t_{0}}\right)$ is defined.
Iteratively conducting these procedures in a gradient-based optimization, we can obtain an optimal solution $\hat{\rmx}_{t_{0}}$ for $\rmx_{t_{0}}$.
We note that the computational complexity to solve the adjoint model from $t_{k}$ to $t_{0}$ has the same order as that of solving the original system (Eq.~\eqref{fwdeq}).
We can obtain an optimal solution with the computational complexity of $O(K_{\text{grad}}C)$, where $K_{\text{grad}}$ is the number of iterations needed in the gradient-based optimization.
Although $K_{\text{grad}}$ is problem-dependent, it is often the case that $K_{\text{grad}}$ is smaller than $n$, if the optimization can be initiated from a good initial guess.
In summary, the 4DVar DA method is efficient when the number of dimensions of $\rmx_{t}$ is large, if only for the purpose of obtaining an optimal solution.

\subsection{Uncertainty quantification}
The 4DVar DA method efficiently provides the optimal solution $\hat{\rmx}_{t_{0}}$ for $\rmx_{t_{0}}$ even when the number of dimensions is large, yet the above-mentioned procedure is not able to quantify the uncertainty in the obtained optimum.
Uncertainty estimation in large-dimensional models has been discussed actively in the field of atmospheric and oceanic DA (e.g., \cite{De05,Bo15}); however, many of the UQ methods have assumed that the given models are linear or can be approximated to be linear.
\revised{In addition, whether the perturbation of the solution grows in time or not can be a crucial matter.}
Since the fault motion model explained in Section~\ref{model} is not only large-dimensional but also non-linear, close attention should be given to the selection of the methods of UQ.
For the purpose of conducting UQ for large-dimensional non-linear models within the 4DVar DA framework, there are several potential methods that are typically classified into three types: (i) Monte-Carlo-based methods, (ii) BFGS-based methods, and (iii) SOA-based methods.
The Monte-Carlo-based methods enable us to construct the variance of the posterior PDF from the perturbations of the results of the 4DVar DA optimization, and the several applications to inversion problems that appear in the field of atmospheric and oceanic DA (e.g., \cite{Ch07,Li14}).
Recently, \cite{Iz21} proposed a method based on the metropolis-adjusted Langevin dynamics (e.g., \cite{Ro78}) with the aim of accelerating sampling from the posterior PDF and applied it to a seismic inversion problem. This method can be interpreted as an advanced version of the conventional Monte-Carlo-based methods.
Although the Monte-Carlo-based methods have the advantage in that they accept any model, it should be noted that the exactness of the obtained results is not always guaranteed because of their stochastic nature, especially in cases of large-dimensional problems.
The BFGS-based methods employ the BFGS method for the gradient-based optimization in the 4DVar DA and construct the approximation of the uncertainty (more precisely, the approximation of the inverse of the second-order derivative matrix of $J$) by the postprocessing of the set of auxiliary vectors obtained in the process of the optimization.
Although the BFGS-based methods are available to any large-dimensional models and have been applied to various scientific fields (e.g., \cite{Fi95,Ge13,Li21}), the accuracy of the approximation obtained by the vanilla BFGS-based method can be poor, depending on the problem properties, even in the linear cases.
Several methods have been proposed to improve the accuracy (e.g., \cite{Bo15,Ni20}), although the problems are limited to being linear.
The SOA-based methods (\cite{It16,It17}) aim at obtaining the inverse of the second-order derivative matrix of $J$ as well as the BFGS-based methods.
Although methods (i) and (ii) pose problems in terms of the accuracy of the uncertainty, the SOA-based methods solve the problem by a combination of a SOA method (\cite{wa92,Di02}) and Krylov subspace method.
Although the implementation of the SOA-based methods can be expensive, they have the advantage in that they enable us to obtain the exact uncertainty up to the round-off error for any non-linear models (\cite{It19}). 
In addition, the SOA-based methods enable us to only compute uncertainty of an element of interest in $\rmx_{t_{0}}$ (more precisely, a column vector of interest in the variance-covariance matrix of $p\left(\rmx_{t_{0}}\mid\Y_{0:k}\right)$).
The wide application range of the SOA-based method can be a motivative reason to apply it to the fault motion model explained in Section~\ref{model}.
In the following section, we explain the details of the SOA-based UQ method.
The SOA-based UQ method starts from considering the Laplace approximation of the posterior PDF $p\left(\rmx_{t_{0}}\mid\Y_{0:k}\right)$ in the neighborhood of the optimal solution.
A Taylor series of the cost function $J$ in the neighborhood of $\rmx_{t_{0}}=\hat{\rmx}_{t_{0}}$ up to the second order leads the approximation to
\begin{equation}\label{lapp}
	p\left(\rmx_{t_{0}}\mid\Y_{0:k}\right) \approx \sqrt{\frac{\text{det}\;\hesse}{\left(2\pi\right)^{\dimx}}}\exp\left[-\frac{1}{2}\left(\rmx_{t_{0}}-\hat{\rmx}_{t_{0}}\right)^{\top}\hesse\left(\rmx_{t_{0}}-\hat{\rmx}_{t_{0}}\right)\right],
\end{equation}
where the matrix $\hesse\in\real^{\dimx\times\dimx}$ is a Hessian matrix evaluated at the optimal solution given by
\begin{equation}\label{hesse}
	\hesse = \sdd{\rmd}{J}{\hat{\rmx}_{t_{0}}}.
\end{equation}
The Laplace approximation says that the variance of each element in $\hat{\rmx}_{t_{0}}$ is given by the diagonal elements of the inverse matrix of Hessian $\hesse^{-1}$.
However, in general, directly evaluating the diagonal elements of $\hesse^{-1}$ is computationally expensive.
Here we estimate the computational complexity to obtain $\hesse^{-1}$ when approximating each of the elements in $\hesse$ via a finite difference of $J$.
The finite difference approximation of each element in $\hesse$ needs a few rounds of computations of $J$, which is proportional to the complexity $C$ to solve Eq.~\eqref{fwdeq}.
This means that we need $O(n^2 C)$ computational complexity to obtain all of the elements in $\hesse$.
Additionally, taking an inverse of $\hesse$ needs the computations proportional to $O(n^3)$ since $\hesse$ is a dense matrix in general.
In summary, the finite difference approximation needs $O(n^2 C)$+$O(n^3)$ computations.
This is unrealistic when $n$ is large; therefore, we employ an alternate method for the computation of $\hesse^{-1}$ proposed by \cite{It16}.
The method can avoid such a large computational complexity since it is dedicated to extracting only the uncertainties of interest.
The key algorithm is the SOA method, which enables the computation of a product of the Hessian $\hesse$ and an arbitrary vector $\rmz$ with the same computational complexity as the one needed to solve Eq.~\eqref{fwdeq}. 
The SOA-based UQ method (\cite{It16}) extracts a specified element in $\hesse^{-1}$ by combining the SOA method and a Krylov subspace method.
The SOA method is composed of solving a set of the tangent linear (TL) model and the SOA model, which are derived from the perturbations of Eqs.~\eqref{fwdeq} and \eqref{adjeq}.
See \cite{wa98,Di02} for the details of derivation. 
The TL model is given by 
\begin{equation}\label{tleq}
	\frac{\rmd \rmtl_{t}}{\rmd t} = \frac{\partial\rmf}{\partial\hat{\rmx}_{t}}\rmtl_{t}, 
\end{equation}
with its inital condition
\begin{equation}
    \rmtl_{t_{0}} = \rmz,
\end{equation}
and the SOA model is given by
\begin{equation}\label{soaeq}
	-\frac{\rmd \rmsoa_{t}}{\rmd t} = \left(\frac{\partial\rmf}{\partial\hat{\rmx}_{t}}\right)^{\top}\rmsoa_{t}
+\left(
\sdd{\partial}{\rmf}{\hat{\rmx}_{t}}
\rmtl_{t}\right)^{\top}\hat{\rmadj}_{t} + \sum_{j=0}^{k}\delta\left(t-t_{j}\right)
\sdd{\partial}{\mathcal{J}_{j}}{\hat{\rmx}_{t_{j}}}
\rmtl_{t_{j}},
\end{equation}
together with its initial and end-point conditions 
\begin{equation}\label{condsoaeq}
\rmsoa_{t_{0}} = \sum_{j=0}^{k}
\sdd{\rmd}{\mathcal{J}_{j}}{\hat{\rmx}_{t_{0}}}
\rmtl_{t_{0}}, \quad  \rmsoa_{t_{f}} = 0,
\end{equation}
where $\hat{\rmx}_{t}$ and $\hat{\rmadj}_{t}$ are the solutions of Eqs.~\eqref{fwdeq} and \eqref{adjeq}, in which $\rmx_{t_{0}}=\hat{\rmx}_{t_{0}}$ is used, and the derivative with respect to $\hat{\rmx}_{t}$ means that with respect to $\rmx_{t}$, to which $\rmx_{t}=\hat{\rmx}_{t}$ is substituted.
Using these models allows us to compute a product of the Hessian and an arbitrary vector, in that, after solving the TL model starting from an initial condition $\rmtl_{t_{0}}=\rmz$, a product of the Hessian $\hesse$ and the vector $\rmz$ is given by using $\rmsoa_{t_{0}}$ as
\begin{equation}\label{hessvec}
	\hesse\rmz = 
	\sdd{\rmd}{I}{\hat{\rmx}_{t_{0}}}
	\rmz + \rmsoa_{t_{0}}.
\end{equation}
The concrete forms of $\sddinline{\partial}{\mathcal{J}_{j}}{\hat{\rmx}_{t_{j}}}$ and $\sddinline{\rmd}{I}{\hat{\rmx}_{t_{0}}}$ will be shown in Section~\ref{App}.
Using the SOA method enables us to access a column vector of the inverse $\hesse^{-1}$ by solving a linear equation 
\begin{equation}\label{lineq}
	\hesse \rmz = \rmb,
\end{equation}
where the vector $\rmb$ is a one-hot vector based on a relevant Krylov subspace method such as the conjugate gradient or conjugate residual methods.
Note that the Hessian-vector computations through Eqs.~\eqref{tleq}--\eqref{hessvec} are  iteratively required to solve Eq.~\eqref{lineq} via the Krylov subspace method.
The uncertainty is given by the square root of the diagonal elements in $\hesse^{-1}$.
The computational complexity needed to solve the set of the TL and SOA models is almost the same as that needed to solve Eq.~\eqref{fwdeq}; therefore, the total computational complexity to obtain an uncertainty through solving Eq.~\eqref{lineq} is $O(K_{\text{Krylov}}C)$, where $K_{\text{Krylov}}$ is the number of iterations needed in the Krylov subspace method until the convergence.
Although $K_{\text{Krylov}}$ is problem-dependent, it is often the case that $K_{\text{Krylov}}$ is smaller than $n$ if the Krylov subspace method can be initiated using a good initial estimate.
This procedure allows us to conduct accurate \revised{UQ} at the highest conceivable efficiency within adjoint-based DA.
The summary of the procedure to obtain an uncertainty is as follows:
\begin{itemize}
    \item[(i)] Obtain $\hat{\rmx}_t$ and $\hat{\rmadj}_t$ by running the 4DVar DA method.
    \item[(ii)] Set the element of the vector $\rmb$ corresponding to the uncertainty element to be one and the others to be zero.
    \item[(iii)] Set the initial $\rmz$.
    \item[(iv)] Solve the linear equation (Eq.~\eqref{lineq}) via a Krylov subspace method, in which the Hessian-vector product computations through Eqs.~\eqref{tleq}--\eqref{hessvec} are required iteratively, starting from the initial $\rmz$.
\end{itemize}

It should be noted that any vector is allowed to be the initial $\rmz$ since Eq.~\eqref{lineq} is a linear equation. In the numerical experiments shown in Section~\ref{App}, we employed a zero vector as the initial $\rmz$.
In addition, note that step (i) can be skipped, and we can reuse $\hat{\rmx}_t$ and $\hat{\rmadj}_t$ when computing other uncertainties since the optimization of the 4DVar DA method and the SOA-based UQ method are split.

\subsection{Numerical integrators to obtain accurate gradients and Hessian-vector products}

\revised{
As is obvious from the formulations in the previous sections, the solution of the forward model (Eq.~\eqref{fwdeq}) is involved in the adjoint (Eq.~\eqref{adjeq}), TL (Eq.~\eqref{tleq}), and SOA (Eq.~\eqref{soaeq}) models, which means that the numerical error in the solution may contaminate the uncertainty.
For this reason, we need to choose a numerical time integrator that provides a sufficiently accurate solution, based on careful consideration of the physical and mathematical properties of the forward model.
Furthermore, the choice of the time integrator for the adjoint model needs to be paid attention to from the perspective of another mathematical property.
}
It is known that there is a symplectic structure between the forward model and the adjoint model; hence a time integrator that does not break this structure after time discretization (see \cite{Sa15} for more details) should be selected. 
As with the numerical time integration of the adjoint model, close attention should also be paid to the choice of the time integrators for the TL and SOA models.
Recently, \cite{It19} reported that the set of four models, i.e., Eqs.~\eqref{fwdeq},\eqref{adjeq},\eqref{tleq}, and \eqref{soaeq}, has a symplectic structure similar to the one involved in the set of Eqs.~\eqref{fwdeq} and \eqref{adjeq}.
For this reason, we need to use the time integrators for the TL and SOA models, which hold the symplectic structure. 
Details on concretely determining the best time integrators can be found in \cite{It19}.
\revised{Using such integrators ensures that the adjoint and SOA models output the exact gradient and Hessian that coincide with the ones obtained by directly substituting the numerical solution of the forward model with the derivatives of $J$.}

\section{Numerical experiments}\label{App}
\subsection{Formulation}
We apply the SOA-based UQ method to the fault motion model explained in Section~\ref{model}.
In this study, the uncertainties of the frictional parameter fields $A_{i}$, $B_{i}$, $L_{i}$ $(i=1,\dots,d)$, and the convergence velocity of the megathrust fault $v_{\text{lock}}$ are of interest.
We assume that $A_{i}$, $B_{i}$, $L_{i}$ $(i=1,\dots,d)$ and $v_{\text{lock}}$ are time-invariant, i.e., their time evolution equations are
\begin{equation}\label{eqpara}
\begin{aligned}
\frac{\rmd A_{i}}{\rmd t} & = 0 \quad \left(i=1,\dots,\dimsim\right) \\
\frac{\rmd \left(A_{i}-B_{i}\right)}{\rmd t} & = 0 \quad \left(i=1,\dots,\dimsim\right) \\
\frac{\rmd L_{i}}{\rmd t} & = 0 \quad \left(i=1,\dots,\dimsim\right) \\
\frac{\rmd v_{\text{lock}}}{\rmd t} & = 0,
\end{aligned}
\end{equation}
respectively.
The set of Eqs.~\eqref{eqth}, \eqref{eqdbk}, and \eqref{eqpara} constitutes the time evolution (Eq.~\eqref{fwdeq}) of the state vector, $\rmx_{t}\in\real^{5d+1}$ defined by
\begin{equation}\label{statevec}
\rmx_{t} =\left( \theta_{1},\dots,\theta_{\dimsim},
				v_{1},\dots,v_{\dimsim}, A_{1},\dots,A_{\dimsim},
		  A_{1}-B_{1},\dots,A_{\dimsim}-B_{\dimsim},
				L_{1},\dots,L_{\dimsim},
				v_{\text{lock}}\right)^{\top}.
\end{equation}
Equation~\eqref{fwdeq} is solved by the RKF45 method as mentioned in Section~\ref{model}, and its adjoint, TL, and SOA models are solved by other types of Runge–Kutta methods that preserve the underlying symplectic structure (see \cite{Sa15,It19} for the details).

This study uses synthetic observational time series data to investigate the influence of data properties (e.g., the signal-to-noise ratio and the period of the time series) on the uncertainties of the frictional parameters.
As the synthetic observational data, we assume a set of snapshots of the velocity fields, i.e.,
\begin{equation}\label{obs}
    \rmv^{\text{obs}}_{\tau} = \obsoperator \rmx_{\tau} + \rmw_{\tau}, \quad \tau\in \mathcal{T}^{\text{obs}}, 
\end{equation}
where $\rmv^{\text{obs}}_{t}\in\real^{d}$ is the observation of the velocity field, $\obsoperator\in\real^{d\times(5d+1)}$ is an observation matrix that extracts elements corresponding to the velocity field from $\rmx_{t}$, and $\mathcal{T}^{\text{obs}}$ is a set of time points at which the observational data exist.
\revised{
We note that Eq.~\eqref{obs} is a bold assumption to make the problem setting simple.
In the real applications, since neither the dimension nor location of the observations matches that of the model in general, the adjustment of the observations and/or the compatible quantities of the model is needed through appropriate interpolation.
}
The observational noise $\rmw_{t}\in\real^{d}$ is assumed to independently and identically follow a normal distribution, i.e.,
\begin{equation}\label{obspdf}
q(\rmw_{t}) = \frac{1}{\left(2\pi\sigma^{2}\right)^{d\slash 2}} \exp\left(-\frac{\rmw_{t}^{\top}\rmw_{t}}{2\sigma^2} \right),
\end{equation}
where $\sigma^{2}$ is the variance.
Equations~\eqref{obs} and \eqref{obspdf} lead the concrete form of $\mathcal{J}_{j}$ and its derivatives needed to implement the adjoint model (Eq.~\eqref{adjeq}) and the SOA model (Eq.~\eqref{soaeq}) as follows:
\begin{equation}
\mathcal{J}_{j} = \frac{d}{2}\log(2\pi\sigma^2) 
+ \frac{1}{2\sigma^2}\left(\rmv^{\text{obs}}_{t_{j}}-\obsoperator \rmx_{t_{j}}\right)^{\top} \left(\rmv^{\text{obs}}_{t_{j}}-\obsoperator \rmx_{t_{j}}\right) \quad (j=0,\dots,k),
\end{equation}
\begin{equation}
\frac{\partial\mathcal{J}_{j}}{\partial\rmx_{t_{j}}} = 
-\frac{1}{\sigma^2}\obsoperator^{\top} \left(\rmv^{\text{obs}}_{t_{j}}-\obsoperator \rmx_{t_{j}}\right) \quad (j=0,\dots,k) ,
\end{equation}
\begin{equation}
\sdd{\partial}{\mathcal{J}_{j}}{\hat{\rmx}_{t_{j}}} = 
\frac{1}{\sigma^2}\obsoperator^{\top}\obsoperator \quad (j=0,\dots,k).
\end{equation}
The prior PDF $p(\rmx_{t_{0}})$ in Eq.~\eqref{post} is assumed to be a normal distribution given by
\begin{equation}\label{priopdf}
p(\rmx_{t_{0}}) = \prod_{i=1}^{5d+1} \frac{1}{\sqrt{2\pi s^{2}_{i}}} \exp\left(-\frac{\left( \left(\rmx_{t_{0}}\right)_{i} -m_{i}\right)^2}{2 s^2_{i}} \right),
\end{equation}
where the mean $m_{i}$ and variance $s_{i}^2$ $(i=1,\dots,5d+1)$ are given element-wise.
From Eq.~\eqref{priopdf}, the explicit forms of $I$ and its derivatives needed in Eqs.~\eqref{gradient} and \eqref{hessvec} are given by
\begin{equation}
I = \sum_{i=1}^{5d+1}\left[ \frac{1}{2}\log(2\pi s_{i}^2) + \frac{\left( \left(\rmx_{t_{0}}\right)_{i} -m_{i}\right)^2}{2 s^2_{i}} \right],
\end{equation}
\begin{equation}
\left(\frac{\rmd I}{\rmd \rmx_{t_{0}}}\right)_{i} = \frac{1}{s^2_{i}}\left( \left(\rmx_{t_{0}}\right)_{i} -m_{i}\right) \quad \left(i=1,\dots,5d+1\right),
\end{equation}
and
\begin{equation}
\left(\sdd{\rmd}{I}{\hat{\rmx}_{t_{0}}}\right)_{ij} =
\frac{1}{s^2_{i}}\delta_{ij} \quad \left(i,j=1,\dots,5d+1\right),
\end{equation}
respectively, where $\delta_{ij}$ is the Kronecker delta.
The uncertainties depend on the parameters related to the observations: the starting time $t_0$ and the time interval $\Delta t$, the number of the observations $k$ (i.e., the end time $t_{k}$ is given by $t_{k}=t_{0}+k\Delta t$), and the parameters in Eqs.~\eqref{obspdf} and \eqref{priopdf}, i.e., $\sigma^2$, $m_{i}$ and $s_{i}^2$.

In this study, we conduct experiments to quantify the uncertainties using a synthetic dataset of the velocity fields, and then investigate how these uncertainties depend on the parameters related to the observations.
We use the solution obtained in Section~\ref{model} as the synthetic data.
Additionally, we assume that the mean $m_{i}$ values are those given in the parameter set used in Section~\ref{model}.
These assumptions make the following analysis easier since the optimal solution $\hat{\rmx}_{t_{0}}$ becomes equivalent to the parameter set used in Section~\ref{model}, and enables us to focus on the relation between the dynamics of slip motion and the uncertainties only.
In addition, we assume $s_{i}$ to be constant values larger than the corresponding elements in $\hat{\rmx}_{t_{0}}$, as shown in Table~2.
\begin{table}\label{sitable}
\caption{Standard deviation $s_{i}$ used in numerical experiments and the state vector $\hat{\rmx}_{t_{0}}$ at which the uncertainties are evaluated. The  vector $\hat{\rmx}_{t_{0}}$ represents the same values as those shown in Table~1.}
\centering
  \begin{tabular}{|c|c|c|} \hline\hline
    & $s_{i}$ & $\hat{\rmx}_{t_{0}}$ \\ \hline
    $A_{i}$ & $1.0\times 10^{3}\;\text{kPa}$ & $1.0\times 10^{2}\;\text{kPa}$ \\ \hline
    $A_{i}-B_{i}$ & $1.0\times 10^{3}\;\text{kPa}$ & \begin{tabular}{rl}$-35 \;\text{kPa}$ & for \; $X^2+Y^2 \le \left( 35\;\text{km}\right)^2$ \\ $70 \;\text{kPa}$ & otherwise \end{tabular}\\ \hline 
    $L_{i}$ & $1.0\times 10^{3}\;\text{mm}$ & $2.2\;\text{mm}$ \\ \hline
    $v_{\text{lock}}$ & $1.0\times 10^2 \;\text{cm}\slash\text{yr}$ & $0.5 \;\text{cm}\slash\text{yr}$ \\ \hline    
    $v_{i}(t_{0})$ & $1.0\times 10^2 \;\text{cm}\slash\text{yr}$ & depends on $t_{0}$ \\ \hline    
    $\theta_{i}(t_{0})$ & $1.0\;\text{yr}$ & depends on $t_{0}$ \\ \hline        
\end{tabular}
\end{table}
The control parameters in the following numerical experiment are the starting time $t_{0}$, the time interval $\Delta t$, the number of observations $k$, and the variance $\sigma^{2}$.
When conducting the UQ based on solving the linear equation (Eq.~\eqref{lineq}) via the conjugate gradient method, some diagonal elements in $\hesse^{-1}$ are sometimes determined to be negative. 
This is because the condition number of the Hessian sometimes becomes too large, depending on the selection of the control parameters.
To obtain the positive diagonals, we switch the solver to the one based on the singular value decomposition when such negative diagonal elements are detected, and then we employ the obtained matrix as a pseudo-inverse of $\hesse$.
In the numerical experiments shown in the next subsection, we mark the result of the pseudo-inverse as “Pseudo” and that obtained by the conjugate gradient method as “Exact”.

\subsection{Time window dependency}\label{twd}
We start by observing how the uncertainties of the frictional parameters depend on the quantities related to the data time window: starting time $t_{0}$, length of time window $\Delta T$, and number of observations $k$.
The numerical experiments in this subsection fix $\sigma$ to be $10^{-3}\;\text{m}\slash\text{yr}$.
\begin{figure}
 \centering
 \includegraphics[width=\preprintfigsize]{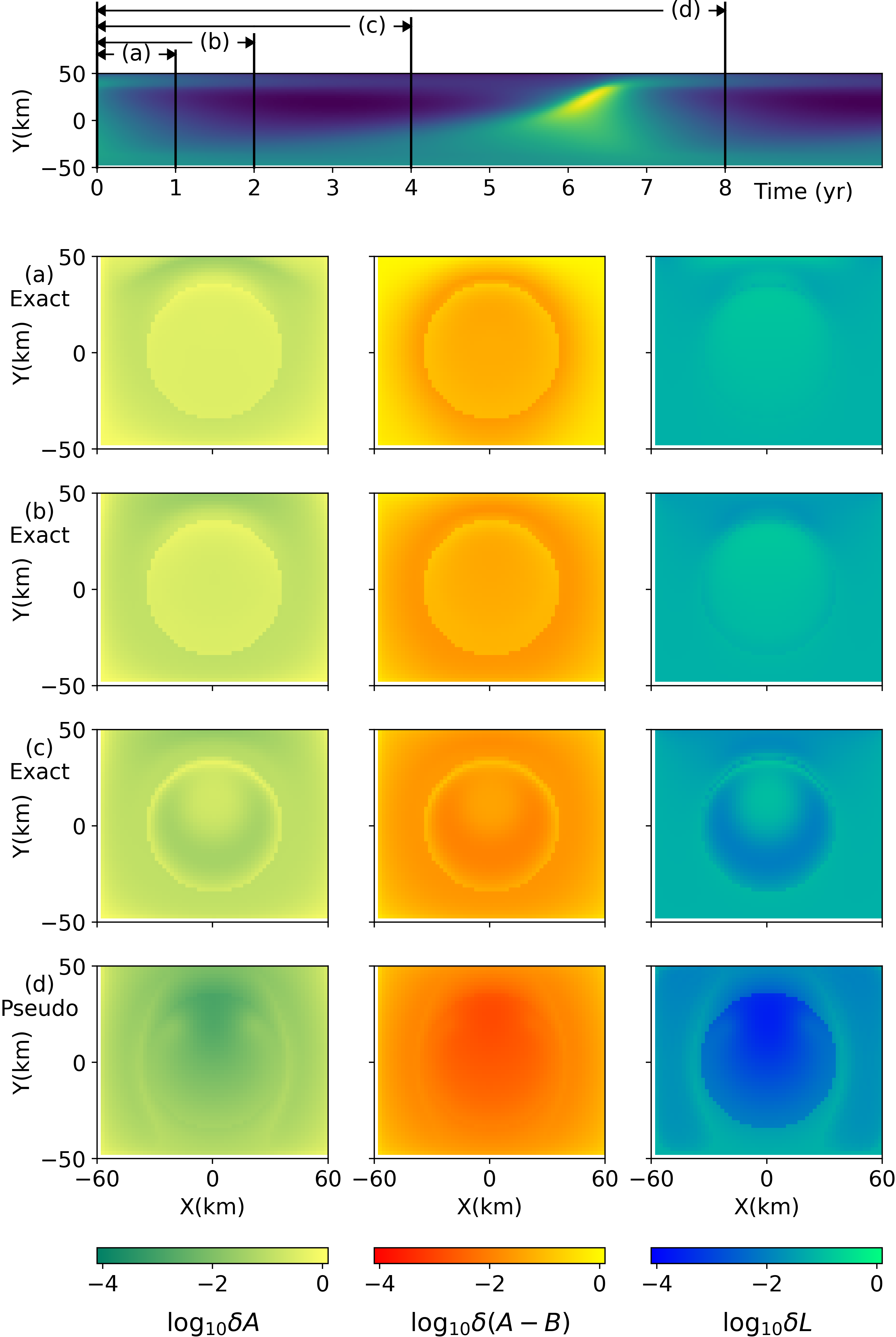}
 \caption{
 Window-length dependency of the uncertainty fields of the frictional parameters.
 The top panel overlays the time windows investigated here on the time evolution of $v$, which is the same as the top panel of Fig.~\ref{evolution}.
 The other panels placed as a matrix show the uncertainty fields of the frictional parameters $A$, $A-B$, and $L$.
 They are normalized by corresponding $s_{i}$ shown in Table~2. The columns indicate each uncertainty field, and the rows indicate the time windows (a)--(d) corresponding to the ones shown in the top panel.}
 \label{Res1}
\end{figure}
\begin{figure}
 \centering
 \includegraphics[width=\preprintfigsize]{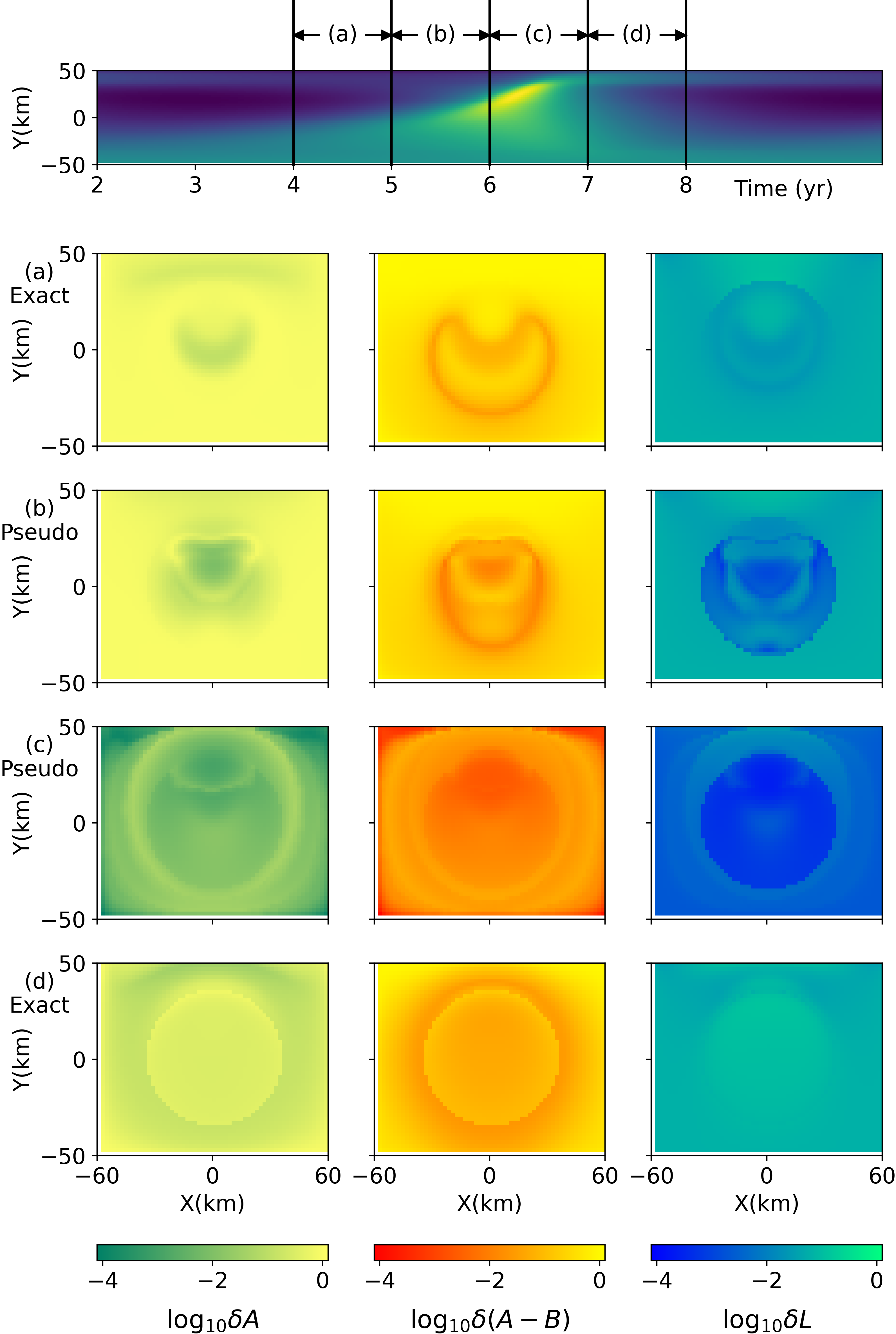}
 \caption{
 Window-position dependency of the uncertainty fields. Each panel shows the same quantity as the corresponding one in Fig.~\ref{Res1}.}
 \label{Res2}
\end{figure}
Figures~\ref{Res1} and \ref{Res2} show the uncertainty fields of the frictional parameters $A$, $A-B$, and $L$.
The uncertainty fields are denoted by $\delta A$, $\delta (A-B)$, and $\delta L$, respectively, and are normalized by their corresponding $s_{i}$ shown in Table~2.
In Fig.~\ref{Res1}, we changed the length of the time window $\Delta T = $ (a) 1, (b) 2, (c) 4, (d) 8 year.
The starting time $t_{0}$ of the time window was fixed to the origin of time of Fig.~\ref{evolution}.
The corresponding number of observations are $k=$ (a) 24, (b) 48, (c) 96, (d) 192, respectively (i.e., the interval between observations is about two weeks).
Contrastingly, in Fig.~\ref{Res2}, we fixed $\Delta T$ to be $1$ year and changed $t_{0}$.
In that case, the number of observations $k$ is $48$ (i.e., the interval between observations is about one week).
Before discussing the details, we emphasize that the uncertainty fields exhibit the information at different scales of richness, fineness of detail, and visual complexity of patterns, depending on the design of the time window.
This is due to our method of enabling the reduction of the computational complexity compared with conventional methods, which do not allow for such reduction.
First, Fig.~\ref{Res1} shows that although the length of the time window works positively to decrease all of the uncertainty fields, the nature of decreasing spatial dependency is different in each uncertainty field.
For the short time window cases ((a) and (b)), the edges around the unstable region determined by the sign of $A-B$ are relatively clear in the uncertainty fields $\delta A$ and $\delta (A-B)$, but misty in the field $\delta L$.
Inversely, for the long time window cases ((c) and (d)), the edges in $\delta A$ and $\delta (A-B)$ are misty, but in $\delta L$, they are clear.
These observations state that, although a long time window is basically recommended to decrease the uncertainty, the sensitivity of $\delta L$ largely depends on the sign of the parameter $A-B$.
It is expected that, with the lengthening of the time window, the uncertainty fields $\delta A$ and $\delta (A-B)$ improve throughout the whole space, but $\delta L$ improves only locally.
This may be due to each frictional parameter's influence on the velocity field, which is the target of the observation data in this study.
The parameters $A$ and $A-B$ affect the time evolution of the velocity fields (Eq.\eqref{eqdbk}) directly; meanwhile, the parameter $L$ affects the fields indirectly through the state variable. 
In other words, the accuracy of $A$ and $B$ are improved proportionally to the increase in information about the velocity, but the improvement of $L$ is slower.
Such a difference in the speed of the influence propagation within the governing equations may account for the difference in sensitivity of the uncertainty with respect to the length of the time window.
Next, we can see from Fig.~\ref{Res2} that although each pattern presents a different level of complexity with respect to understanding its physical origin, the typical magnitude in each uncertainty field becomes smaller (larger) in accordance with the increase (decrease) in the typical magnitude of the velocity field.
In the figure, it can be seen that not only is there a drastic decrease of the entire uncertainty field in (c) that includes the moment when the motion is most active but also a local decrease at the wave front of the velocity field in $\delta (A-B)$ in (a) and (b).
This implies a positive correlation with not only the length of the time window but also between the typical magnitude in the uncertainty field and the activity of seismic motion, meaning that we can obtain an accurate estimation when the activity increases.
Utilizing this property inversely may enable us to predict the activity of seismic motion by monitoring the uncertainty fields.
Throughout the experiments in this subsection, the uncertainty fields were derived from the “pseudo” inverse Hessian if the time window included the moment at which the motion was most active (see (d) in Fig.~\ref{Res1}, and (b) and (c) in Fig.~\ref{Res2}.) 
The inverse Hessian being “Pseudo” means that there is a large magnitude difference between the eigenvalues in the Hessian, thereby providing evidence that the activation of seismic motion induces a drastic decrease in the uncertainty fields.
\subsection{$\sigma$ dependency}
This subsection explains the observations as to how the uncertainty fields depend on the variance $\sigma^2$.
\begin{figure}
 \centering
 \includegraphics[width=\preprintfigsize]{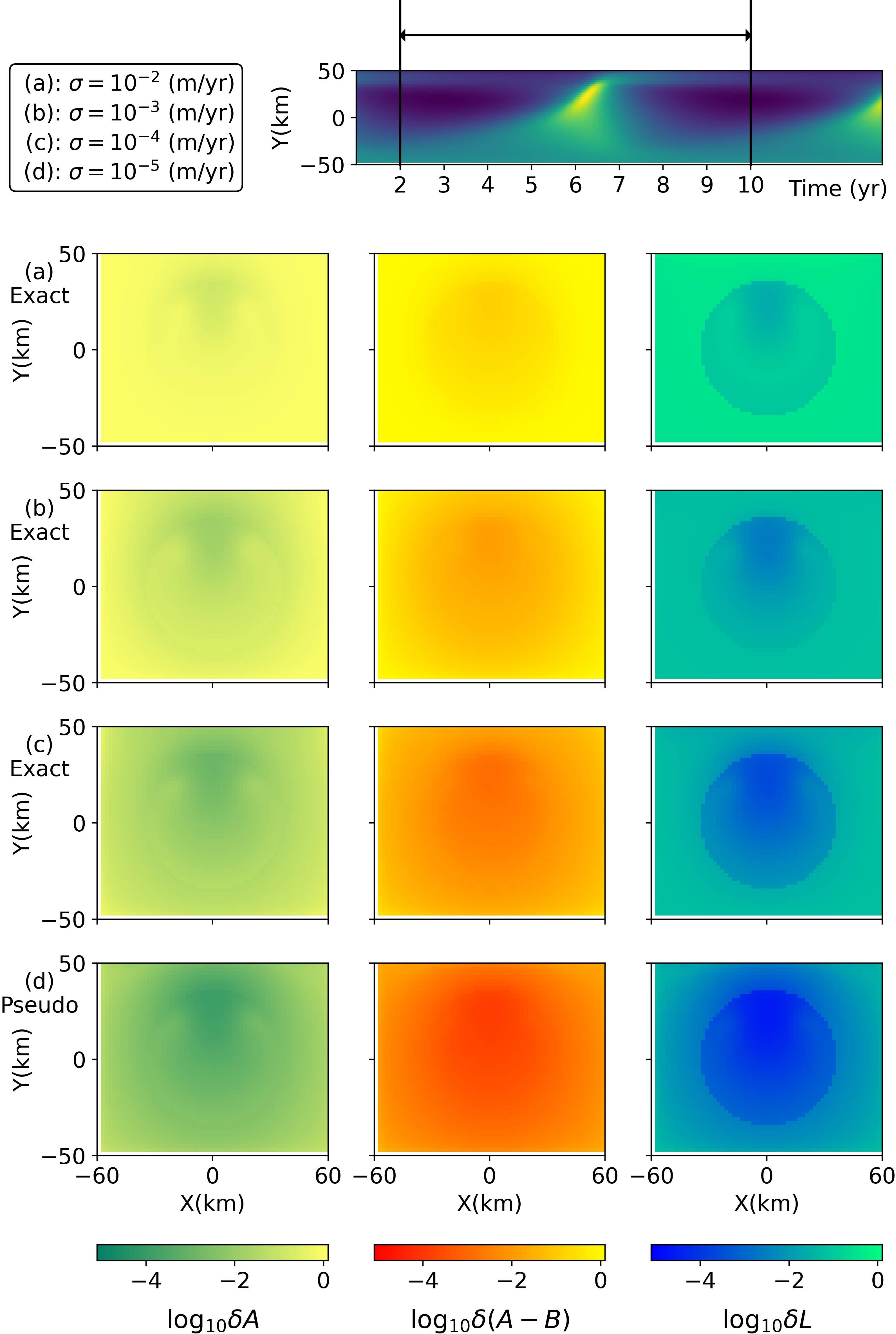}
 \caption{$\sigma$ dependency of the uncertainty fields of the frictional parameters. Each panel shows the same quantity as the corresponding panel in Fig.~\ref{Res1}.
 The common time window shown in the top panel is used among (a)--(d) but different $\sigma$ are used.}
 \label{Res3}
\end{figure}
Figure~\ref{Res3} shows the uncertainty fields obtained by changing $\sigma$ from $10^{-5}\;\text{m}\slash\text{yr}$ to $10^{-2}\;\text{m}\slash\text{yr}$.
The time window in each experiment is fixed to $8$ years and placed as shown in the top panel of the figure, while the number of observations $k$ is fixed to be $8\times 24$ (i.e., the interval of observations is about two weeks.)
The obtained patterns are less complicated compared to those previously obtained, shown in Section~\ref{twd}.
There seems to exist a basic pattern for each quantity; further, the magnitude of the uncertainty decreases monotonically in accordance with the decrease of $\sigma$.
This is easily understandable by considering the contribution of $\sigma$ to the posterior PDF $p\left(\rmx_{t_{0}}\mid\Y\right)$.
From the definition of the Hessian (Eq.~\eqref{hesse}), each element is written by
\begin{equation}\label{hesseelem}
	\hesse_{ij} = \frac{1}{s^{2}_{i}}\delta_{ij} + \frac{1}{2\sigma^2} \frac{\partial^2 R }{\partial (\rmx_{t_{0}})_{i}\partial (\rmx_{t_{0}})_{j}} \quad (i,j = 1,\dots,5d+1),
\end{equation}
where $R$ is the sum of the squared residuals given by
\begin{equation}
R = \sum_{j=0}^{k}\left(\rmv_{t_{j}}^{\text{obs}}-\obsoperator\rmx_{t_{j}}\right)^{\top}\left(\rmv_{t_{j}}^{\text{obs}}-\obsoperator\rmx_{t_{j}}\right).
\end{equation}
Since the fixed time window causes the second-derivative term of $R$ to be a constant matrix, the magnitude of $\sigma$ is a unique factor that is required to determine the final pattern.
Equation~\eqref{hesseelem} tells us that the uncertainty field varies from the uniform pattern to the one determined by $R$ by making the magnitude of $\sigma$ smaller.
The limit $\sigma\rightarrow 0$ lets us expect that the influence of the prior PDF (i.e., the first term on the right-hand side of Eq.~\eqref{hesseelem}) vanishes relative to the second term and allows us to see the uncertainty field depending solely on the observations.
\subsection{Uncertainty of $v_{\text{lock}}$}
Lastly, we observe the uncertainty of the convergence speed of the megathrust fault $v_{\text{lock}}$.
\begin{figure}
 \centering
 \includegraphics[width=\preprintfigsize]{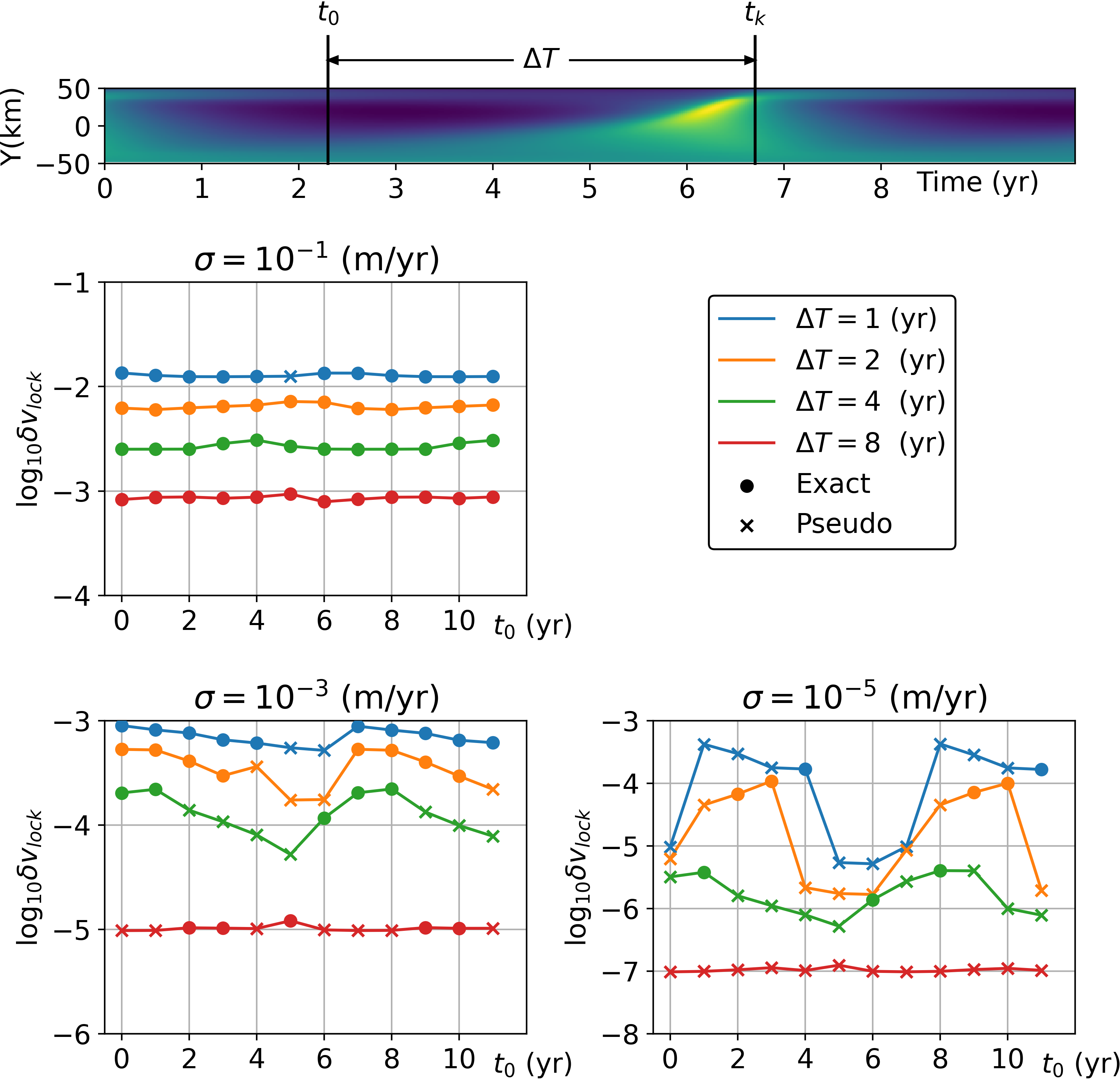}
 \caption{Uncertainty of $v_{\text{lock}}$ with various sets of the variance $\sigma$, the starting time $t_{0}$, and the length of time window $\Delta T$. The color of the line in the middle and two bottom panels indicates the magnitude of $\Delta T$, and the type of point in each panel distinguishes whether the result is “Exact” or “Pseudo”. The uncertainty $\delta v_{\text{lock}}$ in each panel is normalized by $1\;\text{m}\slash\text{yr}$. }
 \label{Res4}
\end{figure}
Figure~\ref{Res4} summarizes the uncertainty of $v_{\text{lock}}$ obtained by the experiments from changing $t_{0}$, $\Delta T$, and $\sigma$.
In this figure, similar behavior with the uncertainty fields shown in the previous sections can be observed, that is, the lengthening of the time window, reduction of $\sigma$, and activation of seismic motion, all of which contribute to the decrease in the uncertainty.
One of the characteristic features observed from Fig.~\ref{Res4} is that the activation of seismic motion does not contribute to decreasing the uncertainty in the large $\sigma$ case (see the middle panel).
In the middle panel, although the blue line ($\Delta T = 1$ yr) contains data that represent both cases, that is, the time windows both including and excluding the moment of seismic motion activation, the overall uncertainty is still almost uniform, just as the other lines depict.
This tells us that (1) we should pay attention in selecting the magnitude of $\sigma$ when conducting the UQ based on real data, and (2) that multiple trial-and-errors to determine the selection of $\sigma$ are essential.
Moreover, from the observation of the bottom panels in Fig.~\ref{Res4}, we find that there is a significant reduction in uncertainty when applying the $\Delta T = 8$ yr time window compared with that when applying the other time windows.
Although the reason for this significant reduction is that the time window includes the activation of seismic motion, it is notable that there is no dependence on the starting time $t_{0}$.
This suggests that whether the time window includes the activation of seismic motion or not is a more crucial factor in reducing the uncertainty than the other possible factors.

\section{Conclusions}\label{conclusion}
This study proposes a method of UQ for inhomogeneous frictional features from the observation of slip motion dynamics. The method has been applied to a seismic model mimicking the slow-slip motion of an LSSE fault along the Bungo Channel in southwestern Japan.
Our UQ method based on the SOA method enables us to obtain high-resolution and accurate uncertainty fields of frictional parameters that have previously been difficult to assess via conventional statistical methods.
The results show that, although the increase in information by lengthening the time window is recommended to achieve the target uncertainty reduction by including the moment at which the seismic motion is most active at the beginning of the time window, the uncertainty reduction can be achieved more effectively.
This provides an important insight into the design of DA and acquisition of actual seismic data, since lengthening the time window and/or integrating too much data may make the computational complexity of a one-time simulation unrealistic.
In addition, our results provide information on the required noise level to obtain the desired uncertainty level, which can provide feedback for future development of observation systems.
Furthermore, since our UQ method provides the uncertainties corresponding to the given model, it suggests whether model improvements are needed or not if extreme data accuracy is required in actual applications to obtain interpretable results.

Regarding the model improvements, as is obvious from the formulation, the DA method used in this study is available under the assumption that the simulation model is the perfect model that covers the dynamics of the objective; thus, it does not apply to the imperfect models that poorly describe the dynamics of the objective in the case of aiming at quantifying the uncertainty purely arising from the observation errors.
%
%
%
%
\revised{
When using such imperfect models for the objective phenomenon, the DA method may overestimate or underestimate the uncertainty, since the imperfect models cannot describe the true state of the phenomenon.
However, the obtained uncertainty is useful even if we have only imperfect models, since it can be the measure of the imperfectness of the model.
The comparison of the uncertainties obtained by applying the DA method to different candidate models with the same data or to the same model with different data helps us decide the direction of model improvement.
}
This study has assumed perfectly-estimated parameters, which are equivalent to the true parameters, to quantify and discuss the uncertainties separately from the effects of randomness in the observational data. In practical situations, we need to pay attention to the possibility of the difference between the optimum solution estimated by the 4DVar DA and the true parameters, which is sourced from the effects of randomness. In addition, depending on the properties of the dataset, the posterior PDF may have a complex multimodality.
Conducting several trials using different datasets and/or different initial guesses of optimization in the 4DVar DA and then comparing their uncertainties is essential in the application to real data in order to obtain reliable results.
In addition, this study employed the prior PDF, being a broad Gaussian with infinite support, to obtain objective results. In the application to the real data, using a prior PDF, by actively integrating the prior knowledge about the possible ranges of the parameters from the perspective of accelerating the optimization in the 4DVar DA, can be useful.

Although our method has reduced the computational complexity compared with conventional methods, to be able to integrate more long-time observations would require even further reduction of computational complexity.
Since it is unavoidable, in principle, that the computational complexity $O(C)$ for a one-time simulation increases by lengthening the time window, it is ideal for reducing our computational complexity $O(nC)$ to the one not proportional to $n$.
One possible method to attain such a reduction may be to employ randomized sampling methods (e.g., \cite{Hu90,Zh18}), which would require accepting a sacrifice in terms of the accuracy of the results.
Randomized sampling methods enable the extraction of the diagonals of the inverse within a constant time that depends on the number of samples.
Although such randomized methods have obtained tremendous success in the fields of machine learning, such as neural networks, applying them directly to the problems represented in this study may be difficult. This is because such methods require the matrix to be sparse enough for a realistic speed improvement; the Hessian matrix in this study is generally not sparse due to the long-range interaction ($K_{ij}$ term) in Eq.~\eqref{eqdbk} that can easily vary the off-diagonals in the Hessian matrix.
Attaining accurate $O(C)$ extraction of the diagonals of the inverse should be the most desired target in future work, yet appropriate solutions remain unknown to the best of our knowledge.

In this study, we successfully quantified and evaluated the high-resolution spatial distribution of slow-slip frictional features of an LSSE fault, which has implications for advancing prediction techniques of megaquake seismic motion through SOA DA-based modeling. The findings are expected to enable the potential for predicting LSSE seismic motion by monitoring the uncertainty fields of frictional parameters.

\begin{acknowledgments}
The authors acknowledge the valuable discussions with the scientists working in the research projects of
JST CREST (grant nos. JPMJCR1761 and JPMJCR1763),
and JSPS KAKENHI Grants-in-Aid for Early-Career Scientists (grant no. 19K14671);
as well as the 
Grant-in-Aid for Challenging Exploratory Research (grant no. 20K21785);
Grant-in-Aid for Scientific Research (S) (grant no. 19H05662);
Grant-in-Aid for Scientific Research (B) (grant nos. 17H01704 and 18H03210);
Grant-in-Aid for Scientific Research (C) (grant no. 22K03542);
the ERI JURP 2020-A-04, 2021-B-01, and 2022-B-06;
and MEXT Project for Seismology toward Research Innovation with Data of Earthquake (STAR-E) grant no. JPJ010217.
The authors also thank two anonymous reviewers for their fruitful comments to improve the manuscript.
\end{acknowledgments}

\section*{DATA AVAILABILITY}
Original data of numerical results associated with this research are available and can be obtained by contacting the corresponding author.

\bibliographystyle{gji2}
\bibliography{refs.bib}

\label{lastpage}

\end{document}